\documentclass[]{pasj02}

\Received{}
\Accepted{}

\usepackage{booktabs}

\usepackage[switch,mathlines]{lineno}
\begin{document}

\title{First Detection of X-ray Polarization and Its Short-term Increase Pre- and Post-Eclipse in HMXB 4U 1700-377}

\author{Kaito Ninoyu\altaffilmark{1}\thefootnote{*}\orcid{0009-0003-0640-2828}}
\email{6223525@ed.tus.ac.jp}
\author{Keisuke Uchiyama\altaffilmark{2}\thefootnote{*}\orcid{0000-0002-7752-9389}}
\email{keisuke.uchiyama@a.riken.jp}
\author{Shinya Yamada\altaffilmark{3}\thefootnote{*}\orcid{0000-0003-4808-893X}}
\email{syamada@rikkyo.ac.jp}
\author{Ryota Hayakawa\altaffilmark{3}\orcid{0000-0001-XXXX-XXXX}}
\author{Shunji Kitamoto\altaffilmark{3}\orcid{0000-0001-8948-7983}}
\author{Nao Kominato\altaffilmark{3}\orcid{0000-0001-XXXX-XXXX}}
\author{Takayoshi Kohmura\altaffilmark{1}\orcid{0000-0003-4403-4512}}
\author{Misaki Mizumoto\altaffilmark{4}\orcid{0000-0003-2161-0361}}
\author{Yuusuke Uchida\altaffilmark{1}\orcid{0000-0002-7962-4136}}
\author{Toru Tamagawa\altaffilmark{5,2}\orcid{0000-0002-8801-6263}}
\author{Ryota Tomaru\altaffilmark{6}\orcid{0000-0002-6797-2539}}
\author{Seoru Ito\altaffilmark{1}\orcid{0000-0003-4403-4512}}

\altaffiltext{1}{Department of Physics and Astronomy, Tokyo University of Science, Chiba 278-8510, Japan} 
\altaffiltext{2}{RIKEN Nishina Center, Saitama 351-0198, Japan} 
\altaffiltext{3}{Department of Physics, Rikkyo University, 3-34-1 Nishi Ikebukuro, Toshima-ku, Tokyo 171-8501, Japan}
\altaffiltext{4}{Science Research Education Unit, University of Teacher Education Fukuoka, Fukuoka 811-4192, Japan}
\altaffiltext{5}{RIKEN Pioneering Research Institute, Saitama 351-0198, Japan} 
\altaffiltext{6}{Department of Earth and Space Science, Osaka University, Osaka 560-0043, Japan}



\maketitle

\begin{abstract}
We report the first statistically significant detection of X-ray polarization from the high-mass X-ray binary (HMXB) 4U 1700-377, observed using the Imaging X-ray Polarimetry Explorer (IXPE). A polarization degree exceeding 10\% was detected above 5 keV, placing it among the highest polarization observed in HMXBs to date. The observation was conducted over a full orbital period of the binary system, during which several sporadic and instantaneous flares were detected. We identify a clear correlation between the polarization degree and orbital phase, with the highest polarization occurring just before and after the eclipse, reaching over 20\% for a few tens of ks. These results suggest that the scattering medium responsible for the observed polarization is spatially localized between the compact object and the O-type companion star, likely created by large-scale inhomogeneities in the stellar wind and its interaction with the compact star's emission.
We also explore the roles of disk winds and orbital reflection in the observed polarization variability. While both mechanisms contribute to the polarization, the substantial increase in polarization before and after the eclipse cannot be fully explained by these models alone, suggesting that the involvement of additional factors. The properties of the X-ray polarization observed by IXPE provide new insights into the accretion processes, X-ray emission, and wind structure in 4U 1700-377, advancing our understanding of their complex environments and the nature of the compact objects within.

\end{abstract}
\section{Introduction}

High-Mass X-ray binaries (HMXBs) are binary systems consisting of a compact object -- either a neutron star (NS) or a black hole (BH) -- and a massive companion star. The compact object accretes matter from its companion, producing X-ray emission. The mode of mass transfer in HMXBs is typically through strong stellar winds from an early-type donor, while low-mass X-ray binaries (LMXBs) accrete via Roche-lobe overflow from a $\lesssim 2~M_{\odot}$ companion.
Notably, HMXBs hosting black holes are considered potential progenitors of merging binary black holes (e.g., \cite{Abbott2023-yc}). The formation, evolution, and mass transfer mechanisms in HMXBs are crucial for understanding their connection to compact object mergers and their role in the context of stellar evolution and gravitational-wave astrophysics.

4U 1700-377 is a well-studied, wind-fed HMXB, initially detected by the Uhuru satellite \citep{Jones1973-ml}, located $\sim$1.7 kpc from Earth \citep{Bailer-Jones2018-vc} with the supergiant HD 153919 as the companion star. The orbital period of 3.4 days with an eclipse duration of $\sim$ 1.1 days was observed \citep{Jones1973-ml}. The mass accretion onto the compact object from the companion star's stellar wind $\sim 10^{-5}~M_{\odot}$ yr$^{-1}$ (\cite{Clark2002-qj}; \cite{Hainich-R2020-ck}) results in bright X-ray emission, ranging from (8--45)$\times$$10^{-10}$erg s$^{-1}$ cm$^{-2}$ \citep{Martinez-Chicharro2018-zv}. During the eclipse period, emission lines containing Fe K$\alpha$ are observed, attributed to scattered and reprocessed X-ray emission from the X-ray source in the stellar wind of the companion star \citep{van-der-Meer2005-or}.

The nature of the compact object in 4U 1700-377 remains a subject of debate. Its measured mass, 2.44$\pm$0.27 $M_{\odot}$ \citep{Clark2002-qj}, places it in an intermediate range between a massive neutron star and a low-mass black hole. Since the Tenma satellite reported a 67-sec periodicity \citep{Murakami1984-xq}, no subsequent confirmations have been reported, nor have any X-ray pulses or outbursts been detected. \citet{Brown1996-hg} proposed that the compact object is a black hole, citing the absence of pulsations and the detection of a hard X-ray tail in the spectrum. In contrast, other studies (e.g., \cite{Reynolds1999-aa}; \cite{Seifina2016-ol}) have explained the observed hard X-ray spectrum using models involving an accreting pulsar or dual Comptonized spectral components with lower electron temperatures, supporting the neutron star hypothesis.
A possible cyclotron line at $\sim$16 keV has been reported \citep{Bala2020-ck}, but with low significance and requiring further confirmation.
If 4U 1700-377 is a neutron star, it could be among the most massive ones without detected pulsations. Alternatively, if it is a black hole, it would be the lowest-mass black hole ever discovered.

The Imaging X-ray Polarimetry Explorer (IXPE; \cite{Weisskopf2022-db}) has significantly advanced our understanding of accretion processes in X-ray binaries. Prior to its launch, \citet{Kallman2015-mu} estimated the polarization of five representative high-mass X-ray binaries (HMXBs): Vela X-1, Cen X-3, Cyg X-1, 4U 1700-37, and SMC X-1. The measured polarization degrees (PDs) are approximately $2.4\%$ for Vela X-1 \citep{Forsblom2023-ud}, $6\%$ for Cen X-3 \citep{Tsygankov2022-xu}, $4\%$ for Cyg X-1 \citep{Krawczynski2022-ok}, and $5\%$ for SMC X-1 \citep{Forsblom2024-jd}. While direct comparison among these sources requires caution due to differences in accretion mechanisms as well as dependencies on spin phase and energy, the PD in HMXBs generally appears to be around or slightly below $10\%$. Although Cyg X-3 -- the only known Galactic binary comprising a compact object and a Wolf-Rayet star -- exhibits >20\% energy-independent linear polarization orthogonal to the radio jet,
indicating a collimated outflow \citep{Veledina2024-nr}, it may represent a somewhat peculiar case.

In weakly magnetized neutron star low-mass X-ray binaries (NS-LMXBs), the polarization is primarily influenced by the hard emission component, whereas the soft emission exhibits lower polarization. This trend is observed in both atoll and Z sources \citep{Ursini2024-wo}. However, the unexpectedly high PD detected in 4U 1820-303 -- an ultra-compact LMXB consisting of a neutron star and a helium white dwarf -- remains challenging to explain. Possible contributions include scattering in an extended corona or wind at high inclination \citep{Di-Marco2023-hb}. Similarly, in magnetars, IXPE observations of sources such as 1E 1841-045 have revealed strong energy-dependent polarization, increasing from approximately $15\%$ to $55\%$, a trend that \citet{Rigoselli2024-xd} reported to be consistent with synchrotron or curvature emission models.

Polarization studies of HMXBs offer a unique opportunity to investigate the mechanisms by which X-rays are polarized through interactions within the accretion onto the compact star and the stellar wind.
In this paper, we present the first polarization observations of 4U 1700-377, obtained during one orbit cycle. The source showed rapid increase in X-rays during the observation.
The PD above $\sim$5 keV exceeds 10\%, a value higher than typically observed in other HMXBs, suggesting that 4U 1700--377 may exhibit distinct polarization properties.
The paper is structured as follows: Section 2 details the observations and data reduction, Section 3 presents the analysis,
and Section 4 discusses the implications of our finding and conclusions.
Unless otherwise stated, all errors quoted in the text and tables, as well as error bars in the figures, correspond to the 1$\sigma$ confidence level.

\section{Observation and Data Reduction}

The IXPE satellite, equipped with three detector units (DUs), each paired with a Mirror Module Assembly, is optimally designed for X-ray imaging polarimetry.
The Gas Pixel Detectors (GPDs) onboard each DU measure the polarization of incident X-ray photons in the energy range of 2--8~keV by determining the direction of the electric field of each polarized photon,
as detailed in \citet{Baldini2021-ze}. For a comprehensive understanding of the polarization measurement principles employed by the GPD, readers are referred to \citet{Di-Marco2022-mf}, \citet{Muleri2022-ms}, and \citet{Baldini2022-jo}.

4U 1700-377 was observed from 2025-02-10T07:46 (UTC) for a total of 150 ks.
Our analysis is based on Level-2 data, which had been processed through the instrumental pipeline,
originating as Level-1 data collected by the GPDs, and subsequently processed using the \texttt{ixpeobssim} software \citep{Baldini2022-jo}, version 31.0.3.
Source event extraction was performed from a circular region with a radius of 120$''$ centered on 4U 1700-377,
utilizing the \texttt{ixpeobssim/xpselect} tool. Given the highly brightness of the central region, the background contribution was negligible \citep{Di-Marco2023-dm}.
Even when a background region was extracted using an annulus with an inner radius of 150$''$ and an outer radius of 300$''$ centered on the source, the analysis results remained unaffected.
Therefore, the background is ignored in the subsequent analysis. For the spectral analysis, we used the response matrices presented in the \texttt{CALDB} and ancillary response files generated with \texttt{ixpecalcarf} for source region.

\begin{figure}[h]
  \begin{center}
    \includegraphics[width=\linewidth]{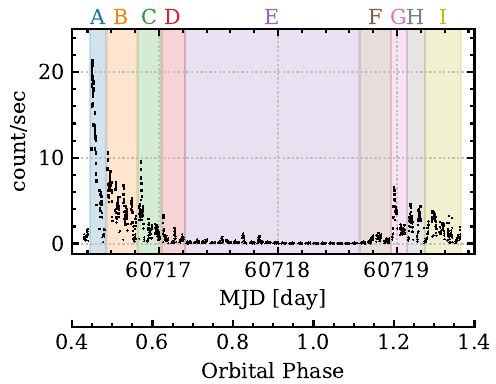}
  \end{center}
  \caption{IXPE light curves and orbital phase. The light curves are shown in the average of the three DUs over the entire energy range. The entire orbital period ($\sim$3.4 days) was covered during the observation.
Alt Text: IXPE light curve averaged over three detectors, with orbital phase show.
  }
  \label{fig:lc}
\end{figure}

The lightcurves of the IXPE are shown in Figure \ref{fig:lc}. The duration of the observation corresponds to one full orbital cycle. We define the orbital phase 0.0 as the superior conjunction of the compact object, when the observer, the companion star, and the compact object are aligned in this order. The X-ray intensity is significantly reduced due to eclipse between orbital phases $\sim$0.7 and $\sim$0.1. X-ray flux exhibits short-timescale variability outside the eclipse. Several types of X-ray flux variability have been studied (e.g., \cite{Doll1987-un}), and they have been interpreted in terms of inhomogeneous stellar wind models, X-ray and wind interactions, or the effects of magnetic fields.

\section{Data Analysis and Results}\label{sec:Data Analysis and Results}

\subsection{Time-averaged X-ray Polarization}\label{subsec:Time-averaged X-ray Polarization}

\begin{figure*}[tb]
  \begin{center}
    \includegraphics[width=0.92\linewidth]{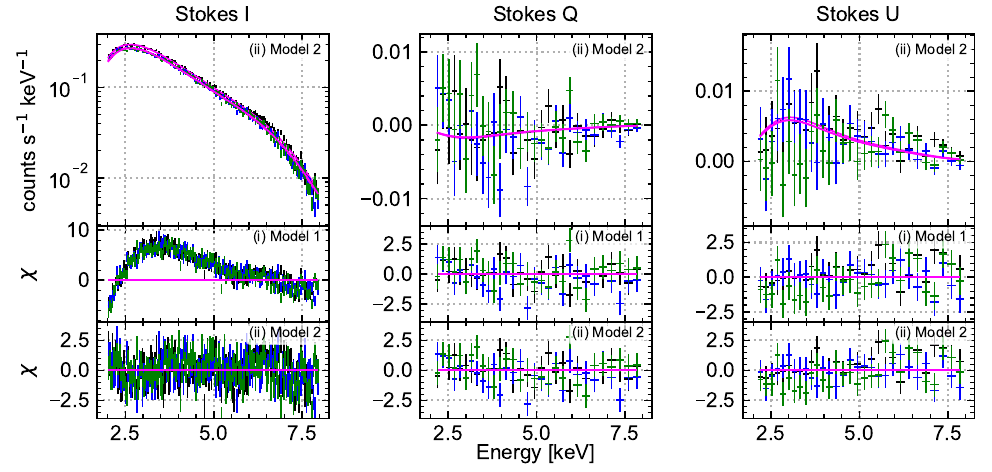}
  \end{center}
  \caption{Time-average spectra of Stokes parameter I (left), Q (center), and U (right). The center and lower panel shows the residuals for (i) Model 1 : a powerlaw model and (ii) Model 2 : a $tbabs \times$ powerlaw model. The best-fit result using Model 2 is shown in magenta overlaid on the top panel. Spectra from the three DUs are plotted in different colors (tight rebinning applied for visual clarity): DU1 in black, DU2 in blue, and DU3 in green.
Alt Text: Time-averaged spectra of Stokes parameters I, Q, and U, with residuals shown for two spectral models.

  }
  \label{fig:iqu}
\end{figure*}

We employed a phenomenological model fitting of Stokes I, Q, and U spectra with \texttt{XSPEC} to understand the time averaged properties.
We first used Model 1: \texttt{const*polconst*powerlaw} model and shows the result in Figure \ref{fig:iqu}.
This model did not reproduce the curvature of the observed spectrum and the fit yielded $\chi^2/$dof = 4857.35/1335.
It is extended to Model 2: \texttt{const*tbabs*polconst*powerlaw}, by including an absorption component.
The results of the spectral fits for Stokes I, Q, and U indicate that the absorbed power-law model provided a good fit ($\chi^2$/dof=1430.84/1334).
The column density is $\sim4\times10^{22}$~cm$^{-2}$, which is similar to those obtained in the past results (e.g., \cite{Seifina2016-ol}).
The Stokes U shows a significantly non-zero signal, whereas the Stokes Q does not.
We find that the PD and polarization angle (PA) shows $7.4\%\pm0.7\%$ and $52^{\circ}.8\pm3^{\circ}.0$ in 2--8~keV.
The flux in the 2--8 keV band derived from the time-averaged data using the best-fit model is $(7.99_{-0.04}^{+0.03}) \times 10^{-10}$ erg cm$^{-2}$ s$^{-1}$. After correcting for interstellar absorption, the unabsorbed flux increases to approximately $9.38 \times 10^{-10}$ erg cm$^{-2}$ s$^{-1}$.
The fitting results are summarized in table \ref{table:fit}.
Assuming a distance of 1.7 kpc, the corresponding luminosity is on the order of $10^{35}$ erg s$^{-1}$, which is consistent with previous studies.

We also analyzed the time-averaged polarization properties using two other methods in addition to polarimetric spectroscopy with \texttt{XSPEC}; \texttt{ixpeobssim/xpbin} PCUBE algorithm \citep{Rankin2022, Kislat2015} and estimation through modulation curve. Polarimetric spectroscopy and the PCUBE algorithm are widely used methods for analyzing polarization data obtained with IXPE. Additionally, we estimated the polarization using the modulation curve, as described in \cite{Ninoyu2024}. Note that the modulation curve with IXPE requires careful interpretation; however, since it represents an important quantity in X-ray polarimetry experiments, we conducted an additional analysis using the modulation curve, as employed in ASTRO-H/SGD \citep{HitomiCollaboration2018}. Figure \ref{fig:fig_pol}(A) shows the time-averaged polarization obtained from the three independent methods in 2--8~keV. These results were consistent with each other, with PD of $\sim$8~\% and PA of $\sim52\%$. Thus, the detection of the first X-ray polarization from this object is considered observationally significant.

\subsection{Energy dependence of the X-ray Polarization}\label{subsec:Energy dependence of the X-ray Polarization}

\begin{figure*}[h]
  \begin{center}
    \includegraphics[width=\linewidth]{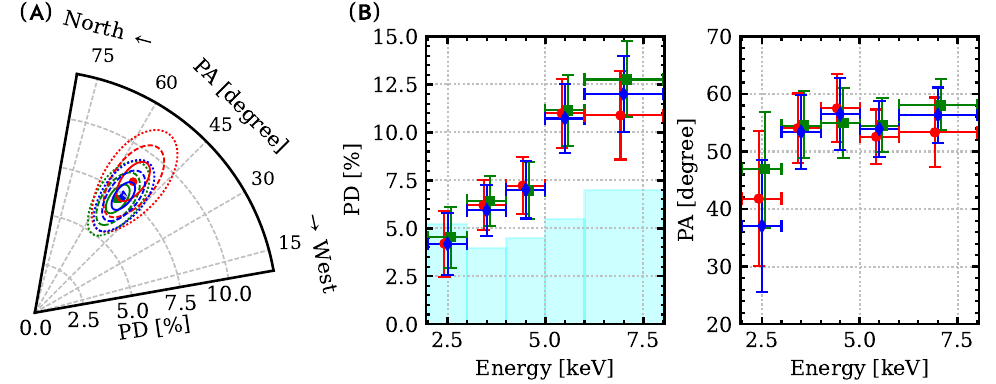}
  \end{center}
  \caption{The obtained PD and PA with the entire observation. (A) The confidence intervals of the PD and PA are shown in polar coordinates. The different colors correspond to different evaluation methods: PCUBE algorithm (green squares), \texttt{XSPEC}-based method (blue diamonds), and modulation curve (red circles) markers. (B)(left) The PD is displayed on the vertical axis, with energy on the horizontal axis, for each method. The cyan hatching indicates the MDP for each energy band. The PD exhibits a clear energy dependence, increasing with energy and reaching $\sim$10\% above 5~keV. (B)(right) The PA is shown on the vertical axis, with energy on the horizontal axis.
Alt Text: Energy-dependent X-ray polarization results from the full observation.
  }
  \label{fig:fig_pol}
\end{figure*}

To investigate the energy dependence of the polarization,
we divided the data into five energy bands and determined the PD and PA for each band.
The spectra in each energy bin were fitted with the Model 2, using the XSPEC-based method, as summarized in table \ref{table:fit}.
For this analysis, all parameters except for the PD and PA were fixed to the values obtained from the fit over the full energy band.
We present the energy dependence of the PD and PA in Figure \ref{fig:fig_pol}(B),
along with the Minimum Detectable Polarization (MDP). The PA remains approximately constant at $\sim55^{\circ}$ above $\sim$3~keV,
whereas the PD increases with energy, exceeding 10\% above 5~keV -- an unusually high value compared to previous results from other HMXBs.

The energy dependence of the observed PD is averaged over the entire observation time and represents a mixture of polarization contributions from both the direct component of the radiation reaching the observer and the scattered component that has undergone at least one scattering event. Therefore, the interpretation not only involves spectral components where the PD varies with energy,
but also suggests that the scattered component has a higher PD, while the direct component has a lower PD.
The varying mixing ratio of these components with energy could account for the observed change in the overall averaged PD.
Given the observational characteristics of this object, including strong absorption and the detection of flares even during eclipses (e.g., \cite{Aftab2019-ez}),
it is plausible to interpret the presence of two distinct radiation components -- scattered and direct -- with the scattered component becoming dominant at higher energies.

\subsection{Time dependence of the X-ray Polarization}\label{subsec:Time dependence of the X-ray Polarization}

\begin{figure}[htb]
  \begin{center}
    \includegraphics[width=\linewidth]{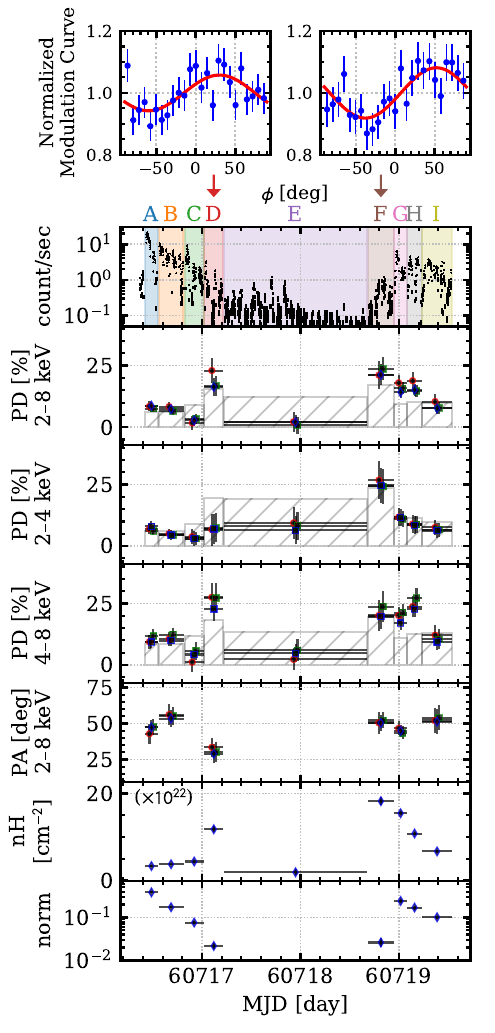}
  \end{center}
  \caption{Summary of the fit results. From top to bottom, the modulation curves in phase D and F, the X-ray count rate, PD (2--8 keV), PD (2--4 keV), PD (4--8 keV), column density, photon index, and normalization of the powerlaw. The hatch marks in the PD (2--8 keV) panel represent the MDP for each time interval. The three colors indicating variability are the same as in Figure \ref{fig:fig_pol}.
  Alt Text:  Summary plot of spectral fit results over time. }
  \label{fig:fig_time_pol}
\end{figure}

To investigate the time variation of such high PDs, the observational data were divided into nine segments (phase A--I), and the PD was examined for each segment. We display the nine segments in Figure \ref{fig:lc} and \ref{fig:fig_time_pol}, each highlighted in a different color. Phase E corresponds to the eclipse phase. X-ray flux variability begins to decrease in Phase D and increases again in Phase F, corresponding to the ingress and egress phases, respectively. We performed spectral fitting with Model 2 for each of the nine segments. The constant factors for DU2 and DU3 were fixed to the values determined from the time-averaged spectrum. As in previous sections, we also estimated the polarization using the PCUBE algorithm and the modulation curve. We find that the PD is low during the eclipse and increases significantly outside of it, particularly just before and after the eclipse. The PD reaches approximately 20\% during both the ingress and egress phases.
The top of figure \ref{fig:fig_time_pol} shows the modulation curves before and after the eclipse, indicating a PD of $\sim$20\%.
The observed modulation amplitude is about 0.06, and the modulation factor is $\sim$0.3 \citep{Di-Marco2022-mf},
yielding a polarization degree of 0.06/0.3 $\simeq$0.2. While caution is generally required in interpreting modulation curves, in this case -- where the source is effectively point-like and the polarization is strong -- the high PD is intuitively evident from the modulation curve.
By testing various time intervals, we confirmed the reproducibility of the high observed PD.

The column density increased toward ingress up to $\sim 10^{23}$cm$^{-2}$, and then gradually decreases after egress.
The duration of phase D and F are $\sim$ 16.8 ks and 22.3 ks, respectively.
The observed orbital phase dependence of the PD strongly suggests that the direct radiation from the compact object does not consistently exhibit high polarization.
Additionally, the high PD observed in phase D and F ($\sim$ 16.8 ks and 22.3 ks, respectively)
is difficult to explain by a reflection component from the companion star, which would be expected to increase when the companion and the compact star are aligned from the observer's perspective.

\section{Discussion and Summary}

We report for the first time the X-ray polarization measurement of 4U 1700-377 using IXPE and investigate the polarization variability in relation to intensity fluctuations and orbital phases.
A correlation between the PD and X-ray energy was observed, with the PD increasing with energy, reaching values above 10\% for energies above $\sim$5 keV.
We also identify a clear correlation between the PD and orbital phase, with the highest polarization occurring just for $\sim$20 ks before and after the eclipse, reaching over 20\%.
This high PD is among highest observed in HMXBs to date, suggesting that the unique properties of this object contribute to such an exceptional result.
Regarding the PA, a difference of $\sim$ 10\% was observed between energies above and below 3 keV.
The energy dependence of the PA provides valuable insights, helping to constrain the origin of the polarization.

The PD varies depending on the polarimetric model, ranging from a few percent up to $\sim$10\%. In the case of an optically thick accretion disk dominated by electron scattering, the expected polarization remains a few percent (e.g., \cite{Dovciak2008-np}, \cite{West2023-dx}). In contrast, the Comptonized component exhibits a higher PD with energy dependence. Assuming an optically thin corona with slab geometry, the PD can reach 10\%-20\% (e.g., \cite{Poutanen1996-kd}). However, for large Thomson optical depths, the PD is expected to decrease.
Therefore, the observed PD exceeding 10\% at high energies in 4U 1700-377 may not be easily explained by simple Compton scattering alone.

In the case of 4U 1820-303, a PD exceeding 10\% has been interpreted as arising from a nonstandard coronal geometry or a relativistic outflow \citep{Di-Marco2023-hb}. In an optically thin corona, the PD of the scattered component can be high, but the presence of unscattered radiation in the observed emission reduces the total PD. While a strong reflection component could contribute to a high PD, its typical value is about 20\% (e.g., \cite{Matt1993-wg}). To explain the observed PD exceeding 10\% above 7 keV, a significant fraction of the direct emission must be obscured.

\begin{table*}[!ht]
\caption{Fitting results of the absorbed powerlaw model, for the entire time and the orbital phases from A to I.}
  \centering
  \begin{tabular}{cccccc}
  \hline\hline
      ~ & time-average & phase A & phaseB & phaseC & phase D  \\ \hline
      PD [\%] & $7.4 \pm 0.7$ & $8.7 \pm 1.5$ & $7.3 \pm 1.5$ & $2.5 \pm 2.0$ & $16.3 \pm 4.0$  \\
      PA [deg] & $52.8 \pm 3.0$ & $47.2 \pm 5.0$ & $53.4 \pm 6.0$ & N/A$^c$ & $28.9 \pm 7.0$  \\
      nH $\times 10^{22}$ [cm$^{-2}$] & $3.985 \pm 0.08$ & $3.35 \pm 0.15$ & $3.77 \pm 0.15$ & $4.39 \pm 0.2$ & $11.76 \pm 0.7$  \\
      $\Gamma$ & $0.292 \pm 0.016$ & $0.835 \pm 0.03$ & $0.795 \pm 0.03$ & $0.582 \pm 0.05$ & $0.228 \pm 0.1$  \\
      norm$^a$ & $0.03108 \pm 0.0008$ & $0.38 \pm 0.02$ & $0.173 \pm 0.009$ & $0.075 \pm 0.006$ & $0.0216_{-0.003}^{+0.004}$  \\
      $C_2$$^b$ & $1.0093 \pm 0.004$ & - & - & - & -  \\
      $C_3$$^b$ & $0.9833 \pm 0.004$ & - & - & - & -  \\ \hline
      $\chi^2/dof$ & 1430.84/1334.0 & 678.81/685.0 & 668.23/685.0 & 659.85/685.0 & 745.56/685.0  \\
      \hline\hline
      ~ & phase E & phase F & phase G & phase H & phase I  \\ \hline
      PD [\%] & $2.0_{-2.0}^{+3.0}$ & $21.3 \pm 4.0$ & $14.2 \pm 2.0$ & $15.2 \pm 2.0$ & $7.6 \pm 2.0$  \\
      PA [deg] & N/A$^c$ & $50.3 \pm 6.0$ & $44.8 \pm 4.0$ & N/A$^c$ & $51.2 \pm 9.0$  \\
      nH $\times 10^{22}$ [cm$^{-2}$] & $1.9_{-0.5}^{+0.7}$ & $18.19 \pm 1.0$ & $15.4 \pm 0.4$ & $10.78 \pm 0.4$ & $6.7 \pm 0.3$  \\
      $\Gamma$ & $-2.09_{-0.07}^{+0.08}$ & $0.53_{-0.12}^{+0.13}$ & $0.75 \pm 0.06$ & $0.77 \pm 0.06$ & $0.979_{-0.05}^{+0.06}$  \\
      norm & $9.0_{-1.1}^{+1.5}\times10^{-5}$ & $0.0261_{-0.005}^{+0.007}$ & $0.241_{-0.02}^{+0.03}$ & $0.17_{-0.017}^{+0.019}$ & $0.1011_{-0.009}^{+0.01}$  \\ \hline
      $\chi^2/dof$ & 1154.40/787.0 & 756.88/685.0 & 637.75/685.0 & 703.68/685.0 & 710.36/685.0 \\ \hline\hline
  \label{table:fit}
  \end{tabular}

\begin{small}
\begin{itemize}
\setlength{\parskip}{0cm} %
\setlength{\itemsep}{0cm} %
\item[$^a$] The normalization of the powerlaw model in a unit of photons keV$^{-1}$ cm$^{-2}$ s$^{-1}$ at 1 keV.
\item[$^b$] The constant factors for the different DUs. These obtained with time-averaged spectra are used in the fits of each phase.
\item[$^c$] The value at this data point could not be reliably estimated due to slightly elevated statistical uncertainties..
\end{itemize}
\end{small}
\end{table*}

4U 1700-377 likely exhibits a high intensity of flares during eclipses, suggesting that the scattering medium may be spatially extended to the scale of the binary system rather than being confined to the vicinity of the compact object. While it remains unclear whether this material originates from a common envelope or stellar wind, its spatial distribution could play a key role. Before and after the eclipse, partial obscuration of the central source could enhance the fraction of scattered radiation, leading to a dominance of the scattered component and an increase in PD.

Variable linear polarization has been observed in the optical radiation from the companion star of 4U 1700-377 \citep{Dolan1984-uw}. The observed intrinsic polarization appears to be correlated with the orbital phase of the binary system, showing variations in the polarimetric light curve over the years. Ultraviolet (UV) polarization in 4U 1700-377, observed using the High Speed Photometer (HSP) onboard the Hubble Space Telescope in four bands (216, 237, 277, and 327 nm), shows a wavelength-dependent PD similar to that seen in Cyg X-1 and 4U 0900-40 \citep{Wolinski1996-pr}. This similarity suggests Rayleigh scattering as the dominant mechanism, likely originating from a gas stream associated with the primary star. The reported PA at orbital phase 0.63 range from 20$^\circ$ to 50$^\circ$ -- except at 277 nm, where the polarization degree is notably lower. At the other wavelengths (216, 237, and 327 nm), the PA is more comparable to that seen in X-rays, indicating that the companion star in 4U 1700-377 may share scattering and wind characteristics with those in the other systems.

The potential contributions of disk winds (\cite{Tomaru2023-bf}, \cite{Nitindala2025-zc}) and orbital variability (\cite{Ahlberg-Varpu2024-dr}, \cite{Rankin2024-na}) to the observed PD in 4U 1700-377 are intriguing, particularly considering the role of accretion disk winds and X-ray radiation reflected from the companion star. Both mechanisms can produce polarization, with disk wind scattering offering a promising explanation for the observed PD at high energies, and orbital variability resulting from reflection processes potentially affecting the PD as well.
However, the high PD observed just before and after the eclipse cannot be explained by the simple sum of these models alone.
Unless the orbital phase and precession (if present) are fortuitously aligned, the observed high PD is unlikely to originate from a physical process occurring close to the compact object.
The observed flare-like variability suggests that additional factors -- such as the enhanced scattering due to partial obscuration or the interplay between multiple emission components -- must be considered.

As a possibility, before and after the eclipse, if there is a dense flow of material, such as an accretion stream or a focused wind considered in Cyg X-1 \citep{Friend1982-qb} or Wind Roch-lobe overflow \citep{Abate-C2013-nz}, the alignment of such dense material, the compact object, and the observer along a straight line could satisfy special geometric conditions that result in high polarization. However, due to significant uncertainties, more detailed observations are needed, including precise spectroscopy with XRISM \citep{Tashiro2020-iz} to clarify the properties of the absorbing material.

\begin{ack}
We deeply appreciate the long development process of the IXPE satellite and the many individuals who have supported it.
\end{ack}

\section*{Funding}
Part of this work was supported by JSPS KAKENHI grant Nos.\ JP24K00672 (S.Y.), 	23K22543 (S.Y.), JP21K13958 (M.M.),
23K25882 and 23H04895 (T.M.), 22H01269, 23K22540 (T.K.), JP19H05609 (K.U. \& T.T.) and Yamada Science Foundation (M.M.).

\section*{Data availability}
This research used data products provided by the IXPE Team (MSFC, SSDC, INAF, and INFN) and distributed with additional software tools by the High-Energy Astrophysics Science Archive Research Center (HEASARC), at NASA Goddard Space Flight Center (GSFC).

\bibliographystyle{aasjournal}
\bibliography{4U1700-37_cvt,Reference}
\end{document}